# How friction and adhesion affect the mechanics of deep penetration in soft solids

Stefano Fregonese [a], Mattia Bacca [a*]

[a]*Mechanical Engineering Department, institute of Applied Mathematics, School of Biomedical Engineering, University of British Columbia, Vancouver BC V6T1Z4, Canada*

[*]Corresponding author. *E-mail address*: mbacca@mech.ubc.ca

**Abstract**
The mechanics of puncture and soft solid penetration is commonly explored with the assumption of frictionless contact between the needle (penetrator) and the specimen. This leads to the hypothesis of a constant penetration force. Experimental observations, however, report a linear increment of penetration force with needle tip depth. This force increment is due to friction and adhesion, and this paper provides its correlation with the properties of the cut material. Specifically, the force-depth slope depends on the rigidity and toughness of the soft material, the radius of the penetrator and the interfacial properties (friction and adhesion) between the two. We observe that adhesion prevails at relatively low toughness, while friction is dominant at high toughness. Finally, we compare our results with experiments and observe good agreement. Our model provides a valuable tool to predict the evolution of penetration force with depth and to measure the friction and adhesion characteristics at the needle-specimen interface from puncture experiments.

**Introduction**
Deep penetration mechanics has been a subject of investigation for several decades thanks to its important engineering and medical applications. It involves a (rigid) penetrator, commonly a needle, that enters deeply into a (soft) material by propagating and opening a crack. This model system is representative of manual and automated medical procedures such as injections [1] blood testing [2], biopsy [3], and other surgical operations [4]. Deep penetration is also observed in several important natural processes in the living kingdom, where tissue penetration is a necessity [5]. Engineering applications involve food processing [6], manufacturing, and (in-situ) material testing [7].

A mechanical description of puncture and deep penetration, in relation to material properties, has been provided via experimentation [8,9,10,11,12,13,14,15,16] and theory [17,18]. Experimental observations evidenced the role of friction and adhesion in mediating needle penetration [15,19,20]. However, a quantitative correlation between frictional forces, material properties and needle radius is still missing. Most investigations rely on the hypothesis of frictionless contact and, thanks to this hypothesis, the penetration force results independent of the depth of the needle tip inside the material. However, this is in disagreement with the observed linear increment of force with depth in puncture experiments (Figure 1, right). This linear increment is due to friction and adhesion between the needle and the specimen and should be quantified to accurately describe the process.

Friction is important in both deep penetration and needle extraction, as observed in medical operations like suture [21], and in nature [5]. Some studies [22,23,24] have accounted for frictional forces at the needle-material interface but did not provide a quantitative link between these forces and the properties of the material.

In this paper we propose a simple model to calculate the slope of penetration force versus depth as a function of interfacial properties (friction and adhesion), material properties (toughness and rigidity) and the radius of the penetrator. This model is finally validated against experiments taken from the literature.

**Deep penetration mechanics**

Our model system is sketched in Figure 1a-c. The needle is a rigid cylinder of radius $R$ penetrating into the specimen to a depth $d$ pushed by a penetration force $P$ (Figure 1a). The mechanical work done by the needle to advance by a depth $\delta d$ is

$$\delta w = P\, \delta d \tag{1}$$

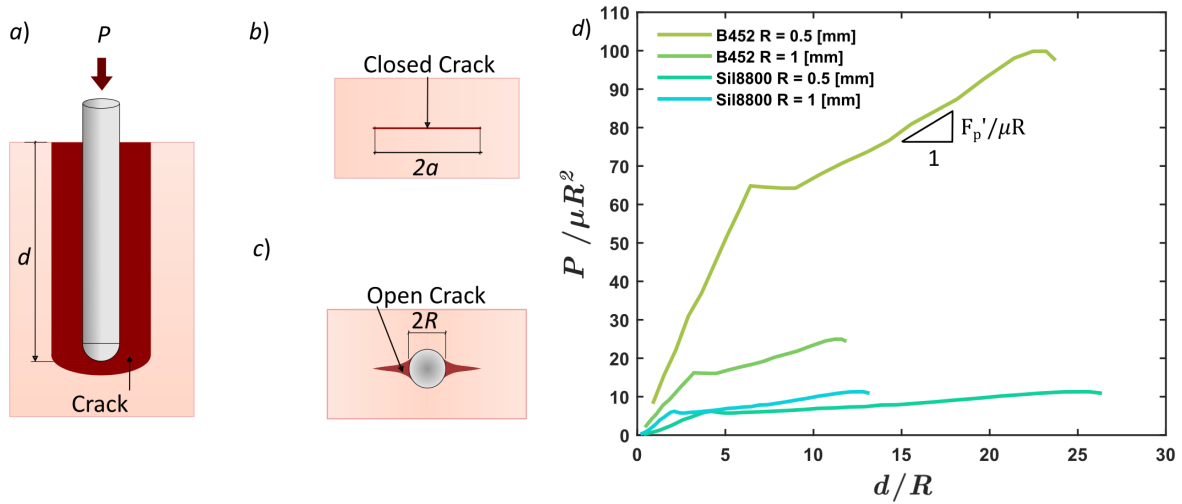

*Figure 1*: Left: schematics of the deep penetration problem; *a)* side view of the penetrator (needle) inserted into the specimen via crack formation and opening, with $d$ the depth of the tip and $P$ the penetration force applied to the back of the needle; *b)* top view of the closed (undeformed) crack; *c)* the crack is opened to host the needle; *d)* Experimental results reporting dimensionless force $P/\mu R^2$ versus dimensionless depth $d/R$ [25].

In frictionless conditions, we have $\delta w = \delta w_0$, which must compensate for the energy required to propagate the crack in the direction of penetration, of undeformed width $2a$ (Figure 1b), by a depth $\delta d$ and to open it so that the needle can slide in (Figure 1c). As described by [17], the required work is

$$\delta w_0 = \left(2 G_c a + \mu R^2\, \hat{h}\right) \delta d \tag{2}$$

where $G_c$ and $\mu$ are the toughness and shear modulus of the material and $\hat{h}$ is a dimensionless function providing the 'spacing energy' to open the crack. By equating Eq. (1) and (2), with $\delta w = \delta w_0$ and $P = P_0$, we obtain the frictionless penetration force

$$P_0 = 2G_c a + \mu R^2 \, \hat{h} \tag{3}$$

As we can see from Eq. (3), $P_0$ is independent of $d$, hence further penetration does not increment the force in frictionless conditions. However, experiments show otherwise, as can be seen in Figure 1d, and we attribute that to friction and adhesion. In this figure, the dimensionless force-displacement plot is truncated where the needle is reaching the bottom of the specimen. Also, the experimental plot reported in Figure 1d is obtained by puncturing rubber with a sharp-tipped (conical) needle, while our model system shows a spherically-tipped needle in Figure 1a. Our model relies on the independence of the penetration force on the needle tip geometry. This is because the crack surface generated by a unit depth increment is $2a$ for any tip having the same radius as the stalk.

**Friction and adhesion**

In the presence of friction, the work required to push the needle by a depth $\delta d$ is

$$\delta w = \delta w_0 + \langle \tau_c \rangle 2\pi R d \, \delta d \tag{4}$$

where $\langle \tau_c \rangle$ is the average contact shear stress $\tau_c$, averaged over the lateral surface $2\pi R d$. In Eq. (4), the contact force $\langle \tau_c \rangle 2\pi R d$ resists sliding and its mechanical work is the second term on the right-hand side.

By equating Eq. (2-4) with (1) we have

$$P = P_0 + \langle \tau_c \rangle 2\pi R d \tag{5}$$

Eq. (5) shows that the penetration force $P$ increments linearly with $d$, as observed in experiments and reported in Figure 1d, with the slope $P' = \partial P / \partial d$, in its dimensionless form

$$\frac{P'}{\mu R} = 2\pi \frac{\langle \tau_c \rangle}{\mu} \tag{6}$$

We describe the resistance to interfacial sliding with

$$\tau_c = \tau_a + \xi \, p \tag{7}$$

where $\tau_a$ is the adhesive shear stress, $\xi$ is the Coulomb friction coefficient, and $p$ is the contact pressure between the specimen and the lateral surface of the needle. While $\tau_a$ and $\xi$ are interfacial properties, $p$ depends on the elastic response of the material and $R$. The adhesion contribution $\tau_a$ accounts for the non-zero sliding resistance in the absence of a contact pressure ($p = 0$). It describes the energetic cost of rupture and reformation of interfacial van der Waals bonds during interfacial sliding and it depends on the chemical affinity between the two materials in contact. When $\tau_a = 0$, Eq. (7) simplifies to Coulomb friction.

By substituting Eq. (7) into (6) one can obtain the force slope. However, the distribution of $\tau_a$ and $p$ is not uniform across the needle-specimen interface, thus requiring detailed definition of the needle-specimen configuration. This depends on the elastic response and toughness of the material, as well as the needle radius.

The incompressible hyper-elastic response of the material is described by the single-term Ogden strain energy density functional [26],

$$\Psi = \frac{2\mu}{\alpha^2}(\lambda_1^\alpha + \lambda_2^\alpha + \lambda_3^\alpha - 3) \qquad (8)$$

For the case of $\alpha = 2$ we obtain Neo Hookean elasticity, while for larger $\alpha$ we obtain a more realistic behavior describing rubber and biological tissue. The Cauchy (true) stress in the material in the principal direction 1 is

$$\sigma_1 = \lambda_1 \frac{\partial \Psi}{\partial \lambda_1} - \Pi \qquad (9)$$

where $\Pi$ is the hydrostatic pressure applied to the material, and $\Psi$ is taken from Eq. (8). Given the interchangeability of $\lambda_1$, $\lambda_2$, and $\lambda_3$ in Eq. (8), courtesy of material isotropy, the principal Cauchy stresses in directions 2 and 3 can be deduced from Eq. (9) by substituting $\lambda_1$ with $\lambda_2$ and $\lambda_3$, respectively.

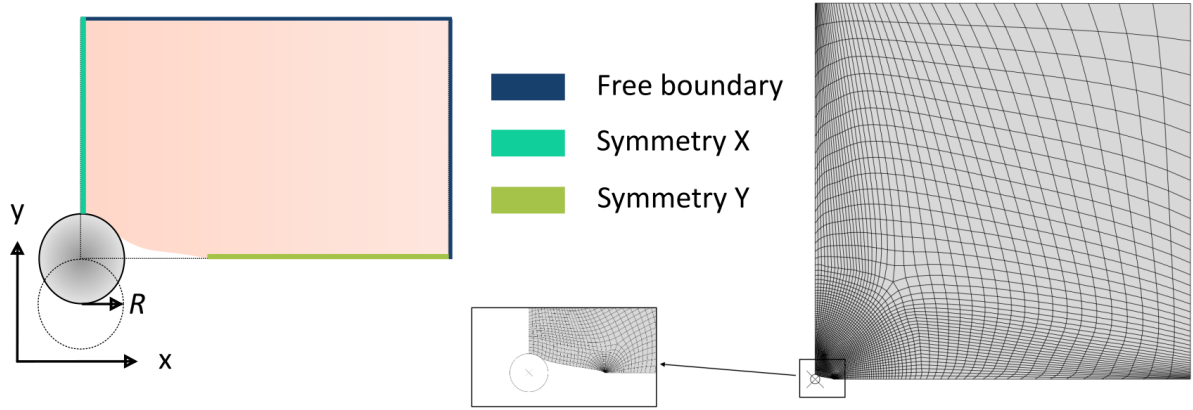

*Figure 2*: Schematics of the finite element model and prescribed boundary conditions. The model represents the cross-section of the needle penetrating the material. Due to symmetry, we consider only a quarter specimen.

To define the needle-specimen configuration described in Figure 1c, we analyze a quarter specimen in plane strain and apply symmetry boundary conditions via finite element analysis (FEA). The model system is sketched in Figure 2. We perform the simulation using the commercial software Abaqus with the mesh depicted in Figure 2. Given the incompressibility of the materials analyzed, we adopted the 'hybrid formulation' considering displacements and pressure as degrees of freedom for each element (element type CPE8H). Due to the stress singularity at the crack tip, we adopted a spider mesh around that region, as shown in Figure 2 (bottom). The FEA computes the J-integral $J$ [27] around the crack tip, *i.e.* the strain energy that would be released by the crack if it advanced by a unit length. This means that $J = \partial U/\partial a$, with $U = \int_\Omega \Psi \, d\Omega$ and $\Omega$ the specimen domain. In this case, $J$, $U$ and $\Omega$ are related to the quarter specimen. The needle-specimen configuration is in equilibrium only if $J = G_c$, following the procedure proposed by [17]. This then provides the correlation between the needle-specimen configuration, $a/R$, and the mechanical behavior of the material as described by $G_c/\mu R$ and $\alpha$ [17,18].

Figure 3 (left) shows the distribution of dimensionless contact pressure $p/\mu$ as a function of the sweep angle $\beta$, for various crack lengths $a/R$ and $\alpha$. $\beta = \pi/2$ identifies the direction of the crack plane. At $\beta = \beta_0$, the contact pressure is zero ($p = 0$) and we define $\beta_0$ as the angle of needle-specimen contact. Because $a/R$ is uniquely defined by $G_c/\mu R$ and $\alpha$ [17,18], we calculate $\langle p \rangle$ and $\beta_0$ numerically, via FEA, as a function of $G_c/\mu R$ and $\alpha$. The pressure distribution reported in this figure consider $a > R$, since $a < R$ gives nearly uniform distribution and $\beta_0 = \pi/2$.

We performed FEA assuming plane strain conditions to calculate $\beta_0$, *i.e.* using plane strain elements, where the needle is pushed against the specimen to reach the configuration in Figure 2 (left). In this case, we assumed frictionless contact and then verified that friction affects $\beta_0$ minimally (within 4% deviation for $\xi = 0.5$ and $a/R = 5$). To calculate $\langle p \rangle$, one could simply integrate the pressure plotted in Figure 3 (left). We used an alternative method and performed a three-dimensional FEA, where the needle is again pushed against the specimen and is then slid against the specimen in the out-of-plane direction (parallel to the needle) to generate a frictional force in that direction. In this case, we describe the interfacial interaction between needle and specimen via Coulomb friction. The frictional force is then divided by the friction coefficient $\xi$ and by the lateral surface of the needle to finally obtain $\langle p \rangle$. To verify our analysis, we compared our results with two values of the friction coefficient, $\xi = 0.01$ and $0.5$, and for $a/R = 5$ and $1.3$. Our results are consistent with a maximum deviation of $0.3\%$. In all the above FEA, to ensure mesh independence, we refined the mesh until obtained consistent results.

Figure 3 (right) plots $\beta_0$ versus $G_c/\mu R$ at various $\alpha$. The numerical results for $\langle p \rangle$ and $\beta_0$ can be fitted to

$$\frac{\langle p \rangle}{\mu} = \frac{\langle p \rangle_0}{\mu} + B_1 \left(\frac{G_c}{\mu R}\right)^{b_1} \tag{10}$$

and

$$\beta_0 = \left(\frac{\pi}{2} - 0.6\right) \exp\left[-B_2 \left(\frac{\mu R}{G_{IC}}\right)^{b_2}\right] + 0.6 \tag{11}$$

with an r-square accuracy of $r^2 \geq 0.9818$, where $\langle p \rangle_0/\mu$, $B_1$, $B_2$, $b_1$ and $b_2$ are fitting coefficients provided in Table 1 as function of $\alpha$. For $G_c/\mu R = 0$, we have the minimum pressure $\langle p \rangle = \langle p \rangle_0$, and angle of contact $\beta_0 = 0.6$.

| $\alpha$ | $\langle p \rangle_0/\mu$ | $B_1$ | $b_1$ | $B_2$ | $b_2$ |
|---|---|---|---|---|---|
| 3 | 0.4483 | 1.603 | 0.4319 | 0.6346 | 0.8253 |
| 5 | 0.4704 | 1.694 | 0.473 | 0.846 | 0.5988 |
| 9 | 0.8613 | 1.434 | 0.844 | 1.267 | 0.4045 |

*Table 1*: Parameters values for Eq. (10) and (11).

**Results and Discussion**

By substituting Eq. (7) into (6), we finally obtain the dimensionless force slope, divided by $\xi$, as

$$\frac{P'}{\mu R \xi} = 4\beta_0 \frac{\tau_a}{\mu \xi} + 2\pi \frac{\langle p \rangle}{\mu} \qquad (12)$$

and plotted in Figure 4 as a function of $G_c/\mu R$ and $\alpha$. In Eq. (12) we considered that $\tau_a$ is applied to the contact region, while the average pressure $\langle p \rangle$ is computed over the total lateral surface of the needle, hence the angles $4\beta_0$ and $2\pi$. In the same equation, $\langle p \rangle/\mu$ and $\beta_0$ are computed from Eq. (10) and (11), and the only interface parameter is $\tau_a/\mu\xi$ representing the ratio of adhesion to friction. When $\tau_a/\mu\xi = 0$ only friction is present, while when $\mu\xi/\tau_a = 0$ we only have adhesion.

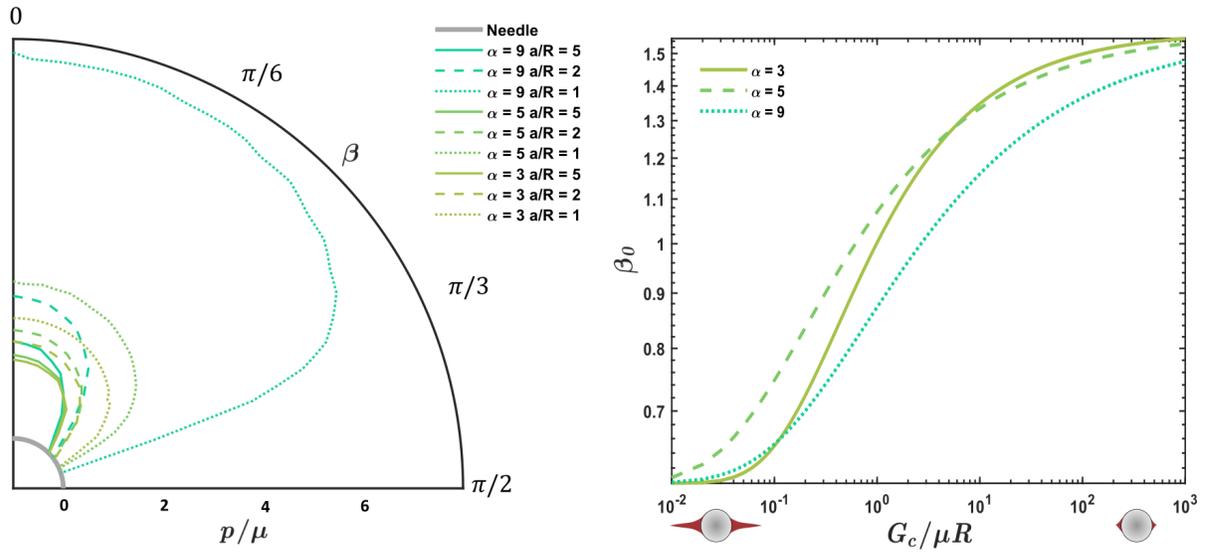

*Figure 3*: Distribution of dimensionless contact pressure $p/\mu$ versus sweep angle $\beta$ (left) (with $\beta = \pi/2$ directed toward the crack plane), for various crack lengths $a/R$ and $\alpha$ (Eq. (8)); Angle of needle-specimen contact $\beta_0$ versus $G_c/\mu R$ (right) for various $\alpha$. At $\beta = \beta_0$, $p = 0$.

Figure 4 also reports a comparison with experimental values taken from [25] and [11]. These experiments did not specify interfacial properties between needle and specimen but only measured the penetration force, and reported material properties and needle radius. "Sil 8800" and "B 425" [25] indicate silicone (with $\alpha = 2.5$, $\mu = 2.7 MPa$, and $G_c = 3.1\ N/mm$) and rubber (with $\alpha = 2.5$, $\mu = 0.4 MPa$, and $G_c = 3.8 N/mm$), respectively. For these materials, the authors reported $\alpha = 3$, $\mu$, $G_c$, $R$, and the $P$-vs-$d$ plot that we used to extract $P'$. Here, we estimated $\xi = 0.03 - 0.05$ and $\tau_a/\mu\xi = 0.5 - 1$. "Gel" [11] indicates a self-assembled triblock copolymer gel with "$\psi = 0.12, 0.16, 0.20, 0.30$" indicating the volumetric fractions of polymer (12, 16, 20, and 30 %). For this material, the authors reported $\mu$, $G_c$, $R$, and the $P$-vs-$d$ plot that we used to extract $P'$ but did not report $\alpha$. [18] estimated $\alpha = 9$ from needle insertion studies on the same experiments. Here, we estimated $\xi = 0.02 - 0.06$, while no estimation of $\tau_a/\mu\xi$ is needed given the small influence of adhesion in this regime.

In the log-log plot of Figure 4 we can observe that the relation between $P'/\mu R\xi$ and $G_c/\mu R$ follows a linear trend at the left and at the right extreme of the plot range (black triangles). This suggests the transition between two asymptotic power-law regimes, giving

$$\frac{P'}{\mu R\xi} \sim \left(\frac{G_c}{\mu R}\right)^q \tag{13}$$

with $q$ the power-law exponent. On the left of Figure 4, for $G_c/\mu R \leq 1$, we have the adhesion-dominated regime and $q$ is controlled by $\tau_a/\mu\xi$ as reported in Figure 5 (left). On the right of Figure 4, for $G_c/\mu R \geq 10^3$, we have the friction-dominated regime and $q$ is controlled by $\alpha$ as reported in Figure 5 (right). In the latter figure, we also plot the fitting functions (dashed line)

$$q = -0.05(\tau_a/\mu\xi)^{0.2535} + 0.31 \tag{14a}$$

for Figure 5-left, and

$$q = 0.0007256\ \alpha^{2.903} + 0.4124 \tag{14b}$$

for Figure 5-right.

Based on the hypothesis of the model, the force-displacement slope $P'$ corresponds also to the slope of the linear reduction in the force required to extract the needle from the specimen. This can be seen in the force-displacement plot during needle retraction in [8], but requires further validation. This model is based on the hypothesis of planar crack formation. However, some experimental results show different crack morphologies, like ring-shaped cracks [14,28-29], conical cracks [15], or more complex morphologies [30]. In these scenarios, our prediction could be inaccurate, and one would have to extend our model to account for complex crack morphologies. We leave this for future work.

**Conclusion**

The proposed model provides the correlation between the penetration force-slope, $P' = \partial P/\partial d$, and the properties of the needle-specimen system. These are toughness, shear modulus and strain-stiffening coefficient of the pierced material, $G_c$, $\mu$ and $\alpha$; the radius of the penetrator (needle) $R$; and the friction coefficient and adhesion shear stress, $\xi$ and $\tau_a$, at the penetrator-material interface. The strain-stiffening coefficient $\alpha$ appears to be relevant only at high material toughness $G_c$, where friction dominates over adhesion. This is due to the high contact pressure exchanged between the needle and the specimen when the crack length $a$ is smaller than the needle radius $R$.

The model can be used to predict $P'$ from the abovementioned properties, if these are known, but can also be used to extract the needle-specimen properties from $P'$ measurements.

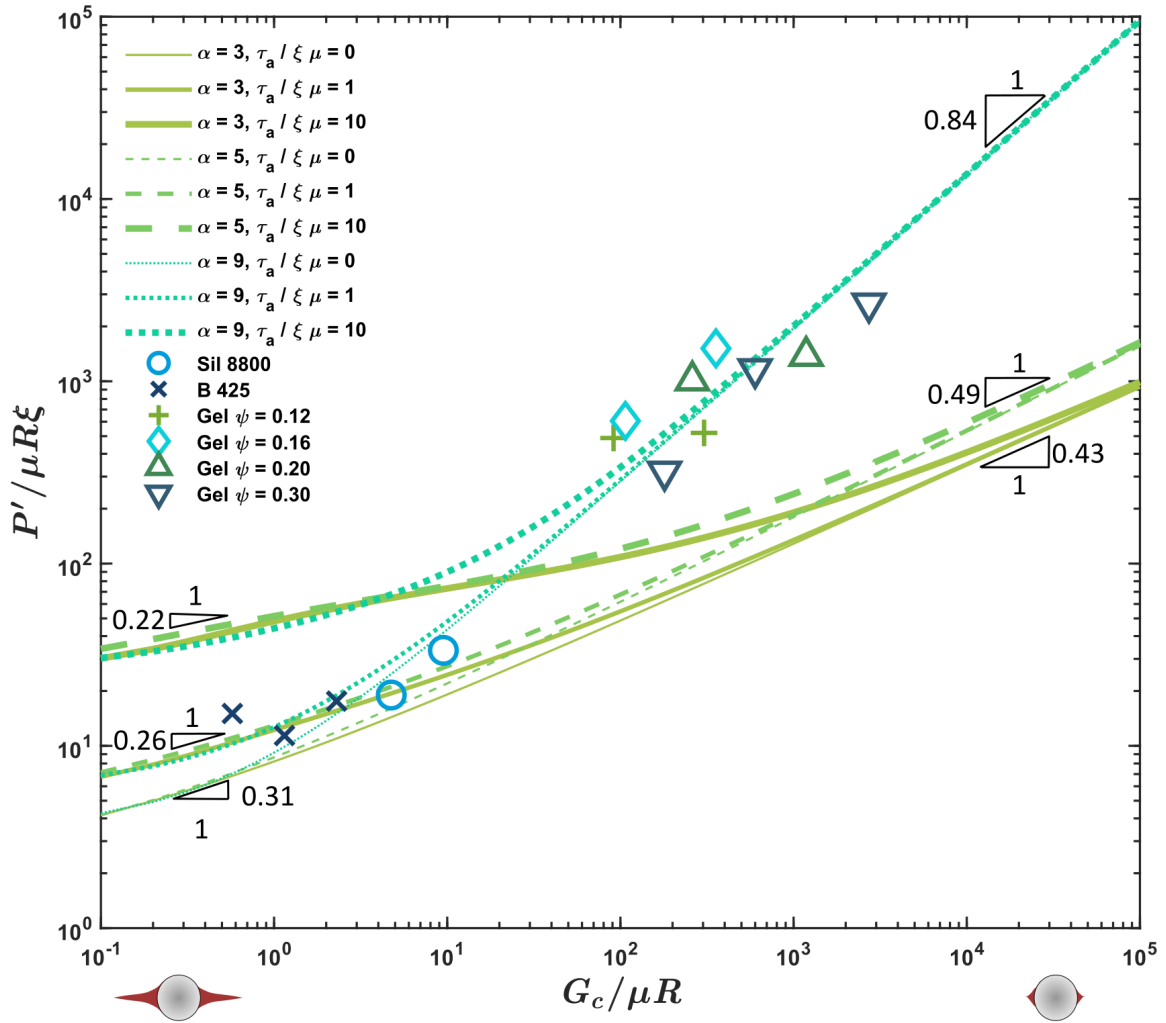

*Figure 4*: Dimensionless force slope $P'/\mu R\xi$ versus $G_c/\mu R$ and $\alpha$: comparison between the numerical predictions generated with the proposed model and experimental observations. Sil 8800 and B 425 are taken from [25]; Gel $\psi = 0.12, 0.16, 0.20, 0.30$ are taken from [11] and indicate a self-assembled triblock copolymer gel with a volume fraction of polymer of 0.12, 0.16, 0.2, and 0.3, respectively.

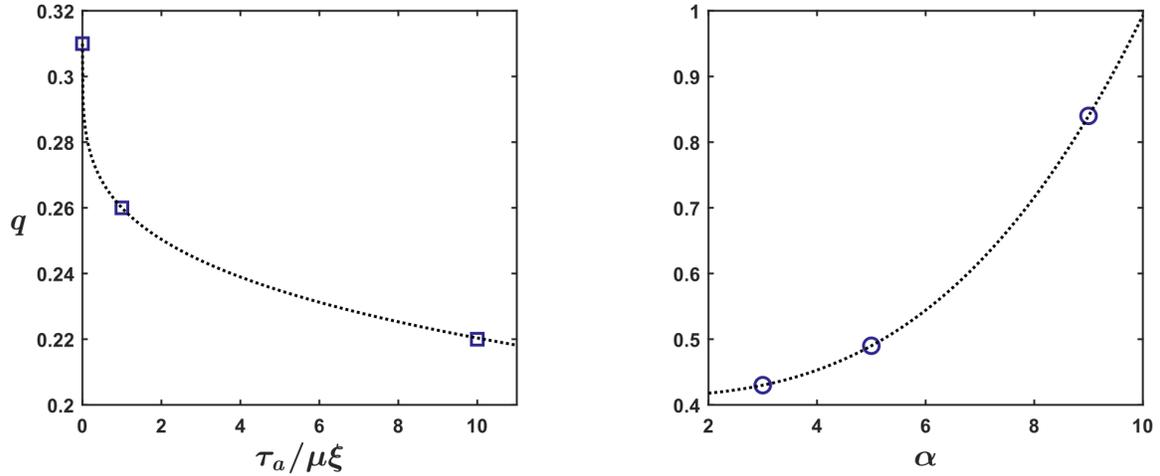

*Figure 5*: Power-law exponent $q$ in Eq. (14) extracted from the asymptotic trends in Figure 4 (black triangles). In the adhesion-dominated regime (left), $G_c/\mu R \leq 1$ and $q$ is function of $\tau_a/\mu\xi$. In the friction-dominated regime (right), $G_c/\mu R > 10^3$ and $q$ is function of $\alpha$.

**Acknowledgments**
The work was supported by the Department of National Defense (DND) of Canada (CFPMN1–026), the Natural Sciences and Engineering Research Council of Canada (NSERC) (RGPIN-2017–04464), and the Human Frontiers in Science Program (RGY0073/2020).

**Bibliography**
[1] S.P. DiMaio, S.E. Salcudean (2003), "Needle insertion modeling and simulation," *IEEE Transactions on Robotics and Automation*, **19**(5):864-875

[2] K. Torossian, M. Ottenio, A.C. Brulez, Y. Lafon, S. Benayoun (2019) "Biomechanics of the medical gesture for a peripheral venous catheter insertion," *Computer Methods in Biomechanics and Biomedical Engineering*, **22**:294–S295

[3] J.Z. Moore, P.W. McLaughlin, A.J. Shih (2012) "Novel needle cutting edge geometry for end-cut biopsy," *Medical Physics*, **39**(1):99-108

[4] Z. Wei, G. Wan, L. Gardi, G. Mills, D. Downey, A. Fenster (2004) "Robot-assisted 3D-TRUS guided prostate brachytherapy: system integration and validation," *Medical Physics*, **31**(3):539-48

[5] W.K. Cho, J.A. Ankrum, D. Guo, S.A. Chester, S. Y. Yang, A. Kashyap, G.A. Campbell, R.J. Wood, R.K. Rijal, R. Karnik, R. Langer, J.M. Karp (2012) "Microstructured barbs on the North American porcupine quill enable easy tissue penetration and difficult removal," *Proceedings of the National Academy of Sciences*, **109**(52):21289-21294

[6] I. Kamyab, S. Chakrabarti, J.G. Williams (1998) "Cutting cheese with wire", *Journal of Materials Science* **33**:2763–2770

[7] Z. Song, S. Cai, (2022), "Needle-induced-fracking in soft solids with crack blunting", *Extreme Mechanics Letter.*

[8] S. Fakhouri, S.B. Hutchens, A.J. Crosby (2015) "Puncture mechanics of soft solids," *Soft Matter*, **11**:4723-4730

[9] C. Gokgol, C. Basdogan, D. Canadinc (2012) "Estimation of fracture toughness of liver tissue: experiments and validation," *Medical Engineering & Physics*, **34**(7):882-91

[10] H. Kataoka, T. Washio, K. Chinzei, K. Mizuhara, C. Simone, A.M. Okamura (2002) "Measurement of the Tip and Friction Force Acting on a Needle during Penetration," T. Dohi, R. Kikinis, *Medical Image Computing and Computer-Assisted Intervention*, MICCAI 2002, Lecture Notes in Computer Science, **2488**:216-223

[11] S. Rattan, A.J. Crosby (2019) "Effect of polymer volume fraction on fracture initiation in soft gels at small length scales," *ACS Macro Letters*, **8**(5):492-498

[12] G. S. Yeh, D.I. Livingston (1961) "The Indentation and Puncture Properties of Rubber Vulcanizates," *Rubber Chemistry and Technology*, 0035-9475

[13] D.I. Livingston, G.S. Yeh, P. Rohall, S.D. Gehman (1961) "Viscoelastic factors in the strength of elastomers under complex stress by a puncture method," *Journal of Applied Polymer Science*, **5**:442-451.

[14] A. Stevenson, K.A. Malek (1994), "On the Puncture Mechanics of Rubber," *Rubber Chemistry and Technology,* **67**(5):743–760.

[15] W.C. Lin, K.J. Otim, J.L. Lenhart, P.J. Cole, K.R. Shull (2009) "Indentation fracture of silicone gels," *Journal of Materials Research* **24:**957–965.

[16] J.N.M. Boots, D. W. te Brake, J.M. Clough, J. Tauber, J. Ruiz-Franco, T.E. Kodger, J. van der Gucht (2022) "Quantifying bond rupture during indentation fracture of soft polymer networks using molecular mechanophores" *Review Materials* **6**:025605.

[17] O.A. Shergold, N.A. Fleck (2004) "Mechanisms of deep penetration of soft solids, with application to the injection and wounding of skin," *Proceedings of the Royal Society of London A*, **460**(2050): 3037-3058

[18] S. Fregonese, M. Bacca (2021) "Piercing soft solids: A mechanical theory for needle insertion," *Journal of the Mechanics and Physics of Solids*, **154**:104497

[19] C.W. Barney, Y. Zheng, S. Wu, S. Cai, A.J. Crosby (2019) "Residual strain effects in needle-induced cavitation," *Soft Matter,* **15**:7390-7397

[20] C.W. Barney, C. Chen, A.J. Crosby (2021) "Deep indentation and puncture of a rigid cylinder inserted into a soft solid," *Soft Matter,* **17**:5574-5580

[21] G. Zhang, T. Ren, S. Zhang, X. Zeng, E. van der Heide (2018) "Study on the tribological behavior of surgical suture interacting with a skin substitute by using a penetration friction apparatus," *Colloids and Surfaces B: Biointerfaces* **162**:228-235

[22] N. Abolhassani, R. Patel, M. Moallem (2007) "Needle insertion into soft tissue: A survey," *Medical Engineering Physics*, **29**(4):413-431

[23] D.J. van Gerwen, J. Dankelman, J.J. van den Dobbelsteen (2012) "Needle–tissue interaction forces a survey of experimental data," *Medical Engineering & Physics*, **34**(6):665–680

[24] G. Ravali, M. Manivannan (2017) "Haptic feedback in needle insertion modeling and simulation," *IEEE Reviews in Biomedical Engineering* **10**:63-77

[25] O.A. Shergold, N.A Fleck (2005) "Experimental investigation into the deep penetration of soft solids by sharp and blunt punches, with application to the piercing of skin," *Journal of Biomechanical Engineering*, **127**(5):838-48

[26] R.W. Ogden (1972) "Large deformation isotropic elasticity – on the correlation of theory and experiment for incompressible rubberlike solids" *Proc. R. Soc. Lond. A* **326**:565–584

[27] R.J. Bucci, P.C. Paris, J.D. Landes, J. Rice (1972) "J integral estimation procedures," *ASTM Special Technical Publication*, 40-69

[28] M. Muthukumar, M.S. Bobji, K.R.Y. Simha (2021) "Needle insertion-induced quasiperiodic cone cracks in hydrogel," *Soft Matter*, **17**:2823-2831

[29] M. Muthukumar, M.S. Bobji, K.R.Y. Simha (2022) "Cone cracks in tissue-mimicking hydrogels during hypodermic needle insertion: the role of water content," *Soft Matter,* **18**:3521-3530.

[30] S.J. Bailey, C.W. Barney, N.J. Sinha, S.V. Pangali, C.J. Hawker, M.E. Helgeson, M.T. Valentine, J.R. de Alaniz (2022), "Rational mechanochemical design of Diels–Alder crosslinked biocompatible hydrogels with enhanced properties," *Materials Horizons,* **9**:1947-1953.